\documentclass[12pt]{article}
\usepackage{epsfig}
\usepackage{amsmath}
\begin{document}

 \title{ Calculation of Positron Channeling Radiation with a Harmonic Potential}
 \author{\sl V.F.Boldyshev, M.G.Shatnev
 \footnote{Corresponding author. E-mail address: mshatnev@yahoo.com.} ,
 \\ Akhiezer Institute for Theoretical Physics of NSC KIPT\\ 1, Akademicheskaya St., Kharkov 61108, Ukraine}
 \date{}
 \maketitle
 \begin{abstract}
  We present a detailed calculation of channeling radiation of planar-channeled positrons from crystal targets in the framework of our approach, which was proposed recently. In contrast to previous calculations of channeling radiation in crystals, our calculation takes into account the interference between different transition amplitudes. The development stemmed from the idea that the amplitude for a given process is the sum of the transition amplitudes for each transition to lower state of transverse energy with the same energy differences between bound-bound transitions. It seems that a consistent interpretation is only possible if positrons move in a nearly harmonic planar potential with equidistant energy levels.

{\bf \textbf{Key words}:} Quantum theory of channeling; Radiation
from positrons;  Harmonic potential

\textbf{PACS codes:}  61.85.+p;  41.60.-m

 \end{abstract}

 \section{ Introduction}
Charged particles directed into a crystal approximately parallel
to one of the crystal planes will be planar channeled. For
positively charged particles, such as positrons, the channel is
between the crystal planes. In classical terms, the particle's
momentum forms some small angle $\theta $ with respect to the
crystal plane. This angle must be less than the Lindhard's
critical angle $\theta _p$ for planar channeling to occur.The
motion of the particles then consists of a periodic back-and-forth
reflection of the boundaries of the planar channel. An accelerated
periodic motion of this kind will lead to the emission of
radiation. From a quantum-mechanical viewpoint, the channel is the
source of a one-dimensional potential well for planar channeling
in the direction transverse to the particle's motion, which gives
rise to transversely bound states for the particle. Its
transitions to lower levels are accompanied by the emission of
radiation with frequencies related to the energy differences of
the levels. Theoretical studies of the radiation from
planar-channeled electrons and positrons are due to M.A. Kumakhov
[1,2], N.K. Zhevago [3], A.I. Akhiezer, I.A. Akhiezer and N.F.
Shul'ga [4,6], A.W.  Saenz, H. Uberall and A. Nagl [5], and to
V.A. Bazylev, V.V. Beloshitsky, V.I. Glebov, N.K. Zhevago, M.A.
Kumakhov and Ch. Trikalinos [7]. Experimentally, the channeling
radiation of positrons was observed by the different groups
[8-11], demonstrating strong and sharp peaks in the spectrum. The
purpose of the present work is to calculate the spectral-angular
distribution of the channeling radiation intensity emitted from
positrons in the framework of approach which was proposed recently
[12].
\section{Wave functions and transverse potential}
The following quantum mechanical calculation of planar positron
channeling radiation utilizes the developments of Kumakhov and
Wedell [2], and Zhevago [3]. One begins with the time-independent
Dirac equation for a relativistic particle moving with momentum
$\overrightarrow{p}_{\parallel }=(0,p_y,p_z)$ in a one-dimensional
planar potential V(x) periodic in the x-direction (which is normal
to the channeling planes)
 \begin{equation}
(i\overrightarrow{\alpha }\cdot \overrightarrow{\nabla }+E-\beta
m)\Psi =V(x)\Psi   \label{eq1}
\end{equation}
where $m$ and $E$ are the particle's mass and energy,
$\overrightarrow{\alpha }$ and $\beta $ are the Dirac matrices.
Separating the wave function $\Psi $ into large and small
components,
\begin{equation}
\Psi =\binom{\Psi _a}{\Psi _b}  \label{eq2}
\end{equation}
 and using the standard representation for the Dirac matrices, leads to a Pauli-type equation for the large components,
 \begin{equation}
\overrightarrow{\sigma }\cdot \overrightarrow{\nabla }(E-V(x)+m)^{-1}%
\overrightarrow{\sigma }\cdot \overrightarrow{\nabla }\Psi
_a+(E-V(x)-m)\Psi _a=0.  \label{eq3}
\end{equation}
Since a potential $V(x)$ is independent of $y$ and $z$, the
solution of Eq. (3) is a plane wave in the $yz$ plane
\begin{equation}
\Psi _a\propto \exp (i(p_zz+p_yy))\varphi (x)\chi.   \label{eq4}
\end{equation}
We now take advantage of the fact that the particle energy is
$E\geq 1\,GeV$, which is much larger than the planar potential
energy, which is on the order of $10\,eV$. This allows us to
transform Eq. (3) into a one-dimensional, relativistic Schrodinger
equation for the transverse motion
\begin{equation}
-\frac 1{2E}\frac{d^2\varphi (x)}{dx^2}+V(x)\varphi
(x)=\varepsilon \varphi (x),  \label{eq5}
\end{equation}
where
\begin{equation}
\varepsilon =\frac{E^2-m^2-p_z^2-p_y^2}{2E}.  \label{eq6}
\end{equation}
In a given potential, the latter will assume certain bound-state
eigenvalues $\varepsilon _n<0\,\,(n=0,1,2...),$ with corresponding
eigenfunctions $\varphi _n(x).$ While Eq. (6) then seems also to
lead to a quantization  of $\left| \overrightarrow{p_{\Vert
}}\right| =\sqrt{p_z^2+p_y^2},$ this is not the case in the
approximation described after Eq. (4) where $\varepsilon \ll E$;
here, Eq.(6) shows that to sufficient accuracy,
\begin{equation}
E\approx E_{_{\Vert }}\equiv \sqrt{\overrightarrow{p_{\Vert
}}^2+m^2}. \label{eq7}
\end{equation}
The wave function of Eq.(2) is finally obtained in the form
\begin{equation}
\Psi =\frac NL\binom \chi {-\frac{i\overrightarrow{\sigma }\cdot
\overrightarrow{\nabla }}{E+m}\chi }\exp (i\overrightarrow{p_{\Vert }\cdot }%
\overrightarrow{r_{\Vert }})\varphi _n(x),  \label{eq8}
\end{equation}
where
\begin{equation}
N=\sqrt{\frac{E+m}{2E}},  \label{eq9}
\end{equation}
$L^2$ being the two-dimensional normalization volume for the plane
waves of Eq.(4), and $\chi $ being a two-component spinor which is
$\binom 10$ or $\binom 01$ when the particle spin points in the
$+z$ or in the $-z$ direction in the rest frame, respectively.
Since $V(x)$ is a periodic potential, we may consider the problem
in a single well with harmonic potential $V(x)=V_0\cdot x^2$ which
describes planar channeling of positrons along the (110) plane of
a silicon crystal [2,3]. For this potential, it is well known that
Eq. (5) can be solved, and the corresponding eigenfunctions being
given by
\begin{equation}
\varphi _n(x)=\sqrt[4]{\frac{E\Omega }\pi }\frac
1{\sqrt{2^nn!}}\exp (-E\Omega x^2/2)H_n(\sqrt{E\Omega }x),
\label{eq10}
\end{equation}
where $H_n$ are the Hermite polynomials, $\Omega $ is the
oscillation frequency given by
\begin{equation}
\Omega =\frac 2{d_p}\sqrt{\frac{2V_0}E},  \label{eq11}
\end{equation}
where $d_p$ is the distance between planes in the corresponding
units, and $V_0$ is the depth of the potential well. The
corresponding eigenvalues $\varepsilon _n$ being given by
\begin{equation}
\varepsilon _n=\Omega (n+1/2).  \label{eq12}
\end{equation}
\section{Matrix element and intensity of channeling radiation}
The crystal plane which channels the positrons is the $yz$ plane,
and the longitudinal component $\overrightarrow{p_{\Vert }}$ of
the positron momentum is taken parallel to the $z$-axis
$(p_y^{}=0)$. The emitted photon $\overrightarrow{k}$ makes an
angle $\theta $ with $\overrightarrow{p_{\Vert }}$, and the
azimuth between the emission $(\overrightarrow{k}\cdot
\overrightarrow{p_{\Vert }})$ plane and the crystal plane is
$\varphi $. Following Kumakhov [2], we can choose two linear
polarization vectors of the photon in the next
way:$\overrightarrow{\varepsilon }_1$ being normal to the emission
plane and $\overrightarrow{\varepsilon }_2$ lying in the emission
plane. The differential emitted photon intensity for a given
polarization is, by Fermi's golden rule,
\begin{equation}
d^2I_\lambda =2\pi \omega \left| M_{if}\right| ^2\delta (\omega
-\omega
\beta _{\Vert }\cos \theta -\stackrel{\sim }{\omega }_{nn^{^{\prime }}})%
\frac{Vd^3k}{(2\pi )^3}\frac{L^2d^2p_{\Vert }^{^{\prime }}}{(2\pi
)^2}, \label{eq13}
\end{equation}
where $V=L^3$ is the normalization volume, $\lambda =1,2$
indicates the linear photon polarization, $\omega $ is the
frequency of the radiation, $\overrightarrow{k}$ is the photon
momentum, $\stackrel{\sim }{\omega }_{nn^{^{\prime }}}\equiv
\varepsilon _n-\varepsilon _{n^{^{\prime }}}$ is the transition
energy between the two levels $n$ and $n^{^{\prime }}$ in the rest
frame of the positron, and $\overrightarrow{p}_{\Vert }^{^{\prime
}}=(0,p_y^{^{\prime }},p_z^{^{\prime }})$ are the components of
the final positron momentum vector parallel to the crystal planes.
The transition matrix element is given by [13]
\begin{equation}
\left| M_{if}\right| ^2=\frac{2\pi e^2}{\omega V}\left| J_\lambda
\right| ^2, \label{eq14}
\end{equation}
where $e^2=\frac 1{137},$ and
\begin{equation}
J_\lambda =\int \Psi ^{^{\prime }\dagger }\alpha _\lambda e^{-i%
\overrightarrow{k}\cdot \overrightarrow{r}}\Psi d^3r, \label{eq15}
\end{equation}
with $\alpha _\lambda =\overrightarrow{\alpha }\cdot \overrightarrow{\varepsilon }%
_\lambda ^{*}$. The prime indicates the final state. Momentum
conservation
\begin{equation}
\overrightarrow{k}=\overrightarrow{p}_{\Vert
}^{}-\overrightarrow{p}_{\Vert }^{^{\prime }},  \label{eq16}
\end{equation}
and energy conservation,
\begin{equation}
E_{_{\Vert }^{}}+\varepsilon _n=E_{_{\Vert }^{}}^{^{\prime
}}+\varepsilon _{n^{^{\prime }}}+\omega ,  \label{eq17}
\end{equation}
lead to the "Doppler shift" formula for the photon energy $\omega
_{nn^{^{\prime }}}$ corresponding to the transition $n\rightarrow
n_{}^{^{\prime }}$,
\begin{equation}
\omega _{nn^{^{\prime }}}=\frac{\stackrel{\sim }{\omega }_{nn^{^{\prime }}}}{%
1-\beta _{\Vert }\cos \theta },  \label{eq18}
\end{equation}
where $\beta _{\Vert }=\frac{\left| \overrightarrow{p}_{\Vert }^{}\right| }{%
E_{_{\Vert }^{}}}$, in the approximation $\omega _{nn^{^{\prime
}}}\ll E$. This approximation is justified for the case $E\sim
GeV$ and $\stackrel{\sim }{\omega }_{nn^{^{\prime }}}=\varepsilon
_n-\varepsilon
_{n^{^{\prime }}}<30\,eV$ considered here. Since $\frac{\left| \overrightarrow{p}_{\Vert }^{}\right| }{E_{_{\Vert }^{}}}%
\approx (1-\frac 12\gamma ^{-2})$ (where $\gamma =\frac Em$),
$\cos \theta \approx (1-\frac 12\theta ^2)$ and $\varepsilon
_n-\varepsilon _{n^{^{\prime }}}=\Omega (n-n^{^{\prime }})$ Eq.
(18) can be expressed as
\begin{equation}
\omega _{nn^{^{\prime }}} =2\gamma ^2\frac{\Omega (n-n^{^{\prime
}})}{(1+\theta ^2\gamma ^2)}. \label{eq19}
\end{equation}
It follows from Eq. (19) that the radiation of a maximum frequency
\begin{equation}
\omega _{nn^{^{\prime }}}=2\gamma ^2\Omega (n-n^{^{\prime }})
\label{eq20}
\end{equation}
is emitted in the forward direction  (at $\theta =0$). The case
$n-n^{^{\prime }}=1$ corresponds to the peak values of the
experimental channeling radiation spectra [11], being the first
harmonic with the photon energy $\omega =2\gamma ^2\Omega $. As it
follows from Eq.(19), photons emitted via positron transition from
any initial level $n$ to the final level $n-1$ are identical
(i.e., have the same energies for the same emission angles). This
means that the resulting amplitude should be given by an additive
superposition of amplitudes of all such transitions. The positron
state outside the crystal $(z<0)$ is a plane wave, whereas inside
the crystal $(z>0)$, the part of its wave function corresponding
to the transverse motion is a superposition of the harmonic
oscillator eigenvectors. Factors $c_n$ describing transitions from
the initial state to states with the transverse energy levels $n$
can be found using boundary conditions set upon the wave function
at the crystal boundary $(z=0)$. Then, a transition to the closest
lower level $n-1$ occurs with emission of a photon having energy
$\omega $. One may expect that the total amplitude of the
transition from the initial to final state accompanied by the
photon emission is determined by products of the amplitudes $c_n$
and $M_{n,n-1}$. Following the rules of the quantum mechanics, we
express this amplitude as
\begin{equation}
\\A \propto\sum_{n}c_nM_{n,n-1} ,\label{eq21}
\end{equation}
were summation is performed over all the harmonic oscillator
levels. Also we must find an additive superposition of amplitudes
of all transitions $n\rightarrow n-2,\;n-3,...$etc. Taking into
account these considerations, we can write the transition matrix
element in the form
\begin{equation}\label{eq22}
\left| M_{if}\right| ^2=\sum_{j}\left| \sum_{n}c_n
M_{n,\,n-j}\right| ^2,
\end{equation}
where $j=n-n^{^{\prime }}$. According to Eq. (15), we find, using
Eq. (8) for the wave functions, the first-order matrix element
corresponding to the $n\rightarrow n-j$ transverse transition
\begin{equation}
J_\lambda =(2\pi )^2\delta (\overrightarrow{p}_{\Vert }^{}-\overrightarrow{p}%
_{\Vert }^{^{\prime }}-\overrightarrow{k_{\Vert }})NN^{^{\prime
}}\chi
_{}^{^{\prime }*}\overrightarrow{\varepsilon }_\lambda ^{*}(\overrightarrow{A%
}+i[\overrightarrow{B}\overrightarrow{\sigma }])\chi ,
\label{eq23}
\end{equation}
where
\begin{eqnarray}
A_x &=&-iI_{n,n-j}^{(2)}(\frac 1{E+m}+\frac 1{E-\omega +m}),  \label{eq24} \\
\overrightarrow{A}_{\Vert }
&=&I_{n,n-j}^{(1)}\overrightarrow{p}_{\Vert
}^{}(\frac 1{E+m}+\frac 1{E-\omega +m}), \\
B_x &=&iI_{n,n-j}^{(2)}(\frac 1{E-\omega +m}-\frac 1{E+m})+\frac{k_x}{%
E-\omega +m}I_{n,n-j}^{(1)}, \\
\overrightarrow{B}_{\Vert } &=&I_{n,n-j}^{(1)}(\frac{\overrightarrow{p}%
_{\Vert }^{}}{E+m}-\frac{\overrightarrow{p}_{\Vert }^{^{\prime
}}}{E-\omega +m}),
\end{eqnarray}
with
\begin{eqnarray}
I_{n,n-j}^{(1)} &=&\int \exp (-ik_xx)\varphi _{n-j}^{*}(x)\varphi
_n(x)dx,
\label{eq2} \\
I_{n,n-j}^{(2)} &=&\int \exp (-ik_xx)\varphi
_{n-j}^{*}(x)\frac{d\varphi _n(x)}{dx}dx,
\end{eqnarray}
Then we calculate the matrix element and the differential
intensity, and after summing over polarization of emitted photon
and the positron using the formulae
\begin{eqnarray}
\varepsilon _i\varepsilon _k^{*} &\rightarrow &\delta
_{ik}-n_in_k,
\label{eq2} \\
\chi _{}^{*}\chi  &=&\frac{1+\overrightarrow{\sigma }\overrightarrow{%
\varsigma }}2\rightarrow \frac 12, \\
\chi _{}^{^{\prime }*}\chi ^{^{\prime }} &=&\frac{1+\overrightarrow{\sigma }%
\overrightarrow{\varsigma }_{}^{^{\prime }}}2\rightarrow 1,
\end{eqnarray}
where $\overrightarrow{\varsigma }$ and $\overrightarrow{\varsigma
}_{}^{^{\prime }}$ are the unit spin vectors of the initial and
final positron (in the rest frame of the positron), respectively,
and taking into account that $\stackrel{\sim }{\omega
}_{nn^{^{\prime }}}=j\Omega $ according to Eq. (12), we find by
integrating over $d^2p_{\Vert }^{^{\prime }}$
\begin{equation}
\frac{d^2I}{d\omega do}=\frac{e^2\omega ^2}{2\pi }\sum_j\left|
\sum_nc_nM_{n,n-j}\right| ^2\delta (\omega -\omega \beta _{\Vert
}\cos \theta -j\Omega ),  \label{eq33}
\end{equation}
where the factors $c_n$ are, in the case of the parabolic
potential and when the initial positron is a plane wave, given by
\begin{equation}
c_n=\frac{i^n}{\sqrt{2^{n-1}n!}}\sqrt[4]{\frac \pi {E\Omega }}\exp (-\frac{%
p_x^2}{2E\Omega })H_n(\frac{p_x^{}}{\sqrt{E\Omega }}).
\label{eq34}
\end{equation}
\section{Conclusion}
Following Kumakhov and Wedell [2], and Zhevago [3], the
spectral-angular distribution of emitted photons is represented as
\begin{equation}
\frac{d^2I}{d\omega do}\propto \sum_f\left| M_{if}\right| ^2.
\label{eq35}
\end{equation}
The sum entering Eq. (35) is the one over the quantum numbers $f$
of the transverse motion of the particle. Then, the probability of
having a definite transverse energy is taken into account by
multiplying each term of this sum by a corresponding factor. In
our consideration, discrete levels of the transverse motion in the
harmonic oscillator potential refer to the intermediate state of
the particle. Accordingly, the contribution to the intensity due
to transitions, e.g., between the closest levels is determined by
the square of the absolute value of Eq. (21)
\begin{equation}
\frac{d^2I^{(1)}}{d\omega do}\propto \left|
\sum_nc_nM_{n,n-1}\right| ^2. \label{eq2}
\end{equation}
In other words, unlike Ref. [2,3], we get an expression that
contains interference terms mixing amplitudes of photon emission
from different equidistant levels. We would like to note that
dynamics of channeling electron in a crystal differs from that of
the positron case. The transverse potential well for the electron
does not give rise to equidistant energy levels for transverse
particle motion. Therefore, there are no interference
contributions to the photon emission intensity similar to those
present in Eq. (36). In our opinion, this could explain the
greater intensity in case of channeling positron compared to that
for the electron observed in experiment [11]. We think that
studying the problems like the one under consideration in this
paper can help in the development of methods for obtaining
polarized photon beams. Corresponding numerical calculations will
be given in a subsequent publication.

 \end{document}